\documentstyle[aps,epsf]{revtex}

\tighten

\makeatletter
\def\leeq{\mathrel{\mathpalette\gl@align<}}
\def\geeq{\mathrel{\mathpalette\gl@align>}}
\def\gl@align#1#2{\lower.6ex\vbox{\baselineskip\z@skip\lineskip\z@
    \ialign{$\m@th#1\hfil##\hfil$\crcr#2\crcr=\crcr}}}
\makeatother

\makeatletter
\def\lesim{\mathrel{\mathpalette\gl@zzalign<}}
\def\gesim{\mathrel{\mathpalette\gl@zzalign>}}
\def\gl@zzalign#1#2{\lower.6ex\vbox{\baselineskip\z@skip\lineskip\z@
    \ialign{$\m@th#1\hfil##\hfil$\crcr#2\crcr\sim\crcr}}}
\makeatother

\newcommand{\BM}[1]{\mbox{\boldmath{$#1$}}}
\newcommand{\PDEL}[2]{\frac{\partial #1}{\partial #2}}
\newcommand{\WDEL}[2]{\frac{d #1}{d #2}}
\newcommand{\BK}[1]{\left( #1 \right)}
\newcommand{\ABS}[1]{\left| #1 \right|}
\newcommand{\STEPF}[1]{\Theta \left( #1 \right)}
\newcommand{\SIKI}[1]{eq.\,(\ref{#1})}
\newcommand{\SIKIL}[1]{Equation\,(\ref{#1})}

\newcommand{\SECT}[1]{\S\,\ref{#1}}

\begin{document}
%\draft

%%%%%%%%%%%%%%%%%%%%%%%%%%%%%%%%%%%%%%%%%%%%%%%%%%%%%%%%%%%
% insert title
\title{Effects of Dissipation on Quantum Phase Slippage \\ 
in Charge Density Wave Systems}

%%%%%%%%%%%%%%%%%%%%%%%%%%%%%%%%%%%%%%%%%%%%%%%%%%%%%%%%%%%
% repeat the \author\address pair as needed
\author{Hiroyuki~MIYAKE and Hiroshi~MATSUKAWA}
\address{Department of Physics,
%\inst{Department of Physics,
         Faculty of science,
         Osaka University}
\date{\today}
%\recdate{\today}

\maketitle
%%%%%%%%%%%%%%%%%%%%%%%%%%%%%%%%%%%%%%%%%%%%%%%%%%%%%%%%%%%
% insert abstract
\begin{abstract}
%\abst{
 We study the effect of the dissipation on the quantum phase slippage
via the creation of ``vortex ring''
in charge density wave (CDW) systems.
The dissipation is assumed to come from the interaction 
with the normal electron near and inside of the vortex core.
We describe the CDW by extracted macroscopic degrees
of freedom, that is, the CDW phase
and the radius of the ``vortex ring'', assume the ohmic dissipation,
and investigate the effect in the context of semiclassical
approximation.
The obtained results are discussed in comparison with experiments.
It turns out that the effect of such a dissipation 
can be neglected in experiments.
%}
\end{abstract}

%%%%%%%%%%%%%%%%%%%%%%%%%%%%%%%%%%%%%%%%%%%%%%%%%%%%%%%%%%%
% insert suggested PACS numbers in braces on next line
%\pacs{????????}
%\kword
%{
%charge density wave (CDW), macroscopic quantum tunneling, 
%dissipation, Caldeira--Leggett theory, vortex shell
%}

%\begin{document}
\sloppy
\maketitle

% body of paper here
%%%%%%%%%%%%%%%%%%%%%%%%%%%%%%%%%%%%%%%%%%%%%%%%%%%%%%%%%%%
%% \baselineskip=24pt
%%%%%%%%%%%%%%%%%%%%%%%%%%%%%%%%%%%%%%%%%%%%%%%%%%%%%%%%%%%
\section{Introduction}
\label{sec:intro}
%%%%%%%%%%%%%%%%%%%%%%%%%%%%%%%%%%%%%%%%%%%%%%%%%%%%%%%%%%%
 
 The sliding motion of a charge density wave (CDW) is one of the most 
interesting phenomena in low dimensional conductors. 
 Under an electric field smaller than a certain threshold field 
$\varepsilon_{\rm T}$, 
the CDW is pinned by impurities, lattice and so on in the sample, 
and the electric conductivity is small.
Above $\varepsilon_{\rm T}$, the CDW starts sliding motion 
and causes a sharp increase of the electric conductivity, 
the spontaneous current oscillation, which is 
known by the name of the narrow band noise, 
and other various peculiar electric conduction 
phenomena \cite{Gruner}. 
 These phenomena are well described by the classical-mechanical treatment 
of the Fukuyama--Lee--Rice (FLR) model \cite{Fuku76a,FuLe78a,LeRi79a}, 
which treats the CDW as a deformable object under pinning 
potentials \cite{Mats87a}.

 Below $\varepsilon_{\rm T}$, the creep motion of the CDW occurs 
with a help of thermal noise 
%by thermal activation 
in the usual temperature range. Then 
the conductance is given as 
\begin{eqnarray}
 \sigma \propto {\rm e}^{- V_{\rm b} / k_{\rm B}T},
\end{eqnarray}
where $V_{\rm b}$ is the effective barrier height.
  In the lower temperature range, however, the quantum effect is 
expected to be important. 
In fact Zaitsev--Zotov observed in o-${\rm TaS}_3$ a temperature 
independent nonlinear
conduction below $\varepsilon_{\rm T}$ in the temperature range 
below $10$ K, which is considered to be due to 
quantum mechanical tunneling of the CDW \cite{ZaZo93a}.
 The experiment was carried out in the two-probe configuration.
% Zaitsev--Zotov \cite{ZaZo93a} 
He also found that the following dependence of 
the conductance $\sigma$ on the electric field $\varepsilon$,
\begin{eqnarray}
  \sigma \propto {\rm e}^{- \left( \varepsilon_0 / \varepsilon \right)^2} ,
\label{eq:exdep}
\end{eqnarray}
where $\varepsilon_0 \approx  1.1 \times 10^4 [{\rm V/m}]$ 
in the range $3 \times 10^3 [{\rm V/m}] \lesim \varepsilon \lesim 
1 \times 10^4 [{\rm V/m}]$. 
This form is different from the one due to the Zener-tunneling 
\cite{Bard79a} or the creation of kink--antikink pairs \cite{Maki77a}.
 The above nonlinear conduction below $\varepsilon_{\rm T}$ 
is observed only in thin samples.
 
\begin{figure}
\begin{center}
\leavevmode
\epsfxsize=0.70\textwidth
\epsfbox{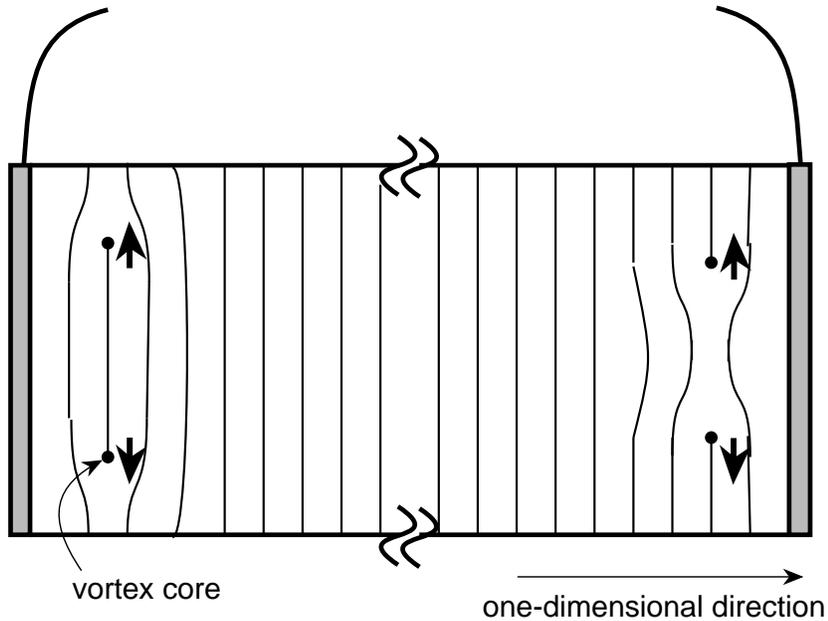}
\end{center}
\caption{The phase configuration in sample. The top of CDW, 
where ${\bf\mit Q}\!\cdot\!{\bf\mit r}+\phi =
2 n \pi$, is shown by solid lines 
in the plane perpendicular to the one--dimensional direction. 
The dots show the vortex cores. Moving vortex core, that is, 
expanding ``vortex ring'', converts the condensed electrons 
to the normal electrons or opposite.}
\label{fig:1}
\end{figure}

 In the framework of the semiclassical theory, the CDW current 
due to the quantum tunneling is related to the decay rate of the 
metastable state. Then the conductance is expressed as
\begin{eqnarray}
  \sigma \propto {\rm e}^{- S_{\rm B} / \hbar} .
\end{eqnarray}
Here $S_{\rm B}$ is the Euclidean action for the ``bounce'' solution, 
which is the least action solution, 
corresponding to the proper tunneling process \cite{Cole77a,CaCo77a}.
 Duan proposed the ``vortex shell'' in the $3+1$ 
dimensional space as 
the ``bounce'' solution, and obtained the same electric field 
dependence as that in \SIKI{eq:exdep} \cite{Duan93a}.
 The vortex shell represents 
the conversion process from the single--particle current 
to the CDW current or opposite by expanding 
the dislocation loop (or ``vortex ring'' in terms of the phase) 
in the perpendicular plane to the one--dimensional direction 
as shown in Fig.\,\ref{fig:1}. The motion of dislocation loop  
causes the phase slippage \cite{Maki86a,Haya98a}.
 The magnitude of $\varepsilon_0$ in \SIKI{eq:exdep} 
that Duan estimated is, however, about $10^3$ times as large as that 
in the experiment.
 This is because he assumed the homogeneous CDW system, 
the quantum nucleation can occur wherever in the sample 
with the same probability, 
and the energy gain due to the creation of a vortex shell comes from 
the interaction of the CDW with the external electric field. 
 Maki \cite{Maki95a} noticed that the experiment was carried out
in the two-probe configuration, and considered the strong pinning by 
the electric probes at both ends of the sample at first. 
 He reset the distance between the probes to the FLR coherence length 
in the case of $3$D weak pinning to get the agreement with the experiment. 
The reason of the replacement is, however, not clear.
%\begin{minipage}[t]{80mm}
%\begin{figure}
%\begin{center}
%\leavevmode
%\epsfxsize=0.80\textwidth
%\epsfbox{fig1.eps}
%\end{center}
%\caption{The phase configuration in sample. The top of CDW, 
%where ${\bf\mit Q}\!\cdot\!{\bf\mit r}+\phi =
%2 n \pi$, is shown by solid lines 
%in the plane perpendicular to the one--dimensional direction. 
%The dots show the vortex cores. Moving vortex core, that is, 
%expanding ``vortex ring'', converts the condensed electrons 
%to the normal electrons or opposite.}
%\label{fig:1}
%\end{figure}
%\ 
%\end{minipage}
%%
%\hspace{\fill}
%%
%\begin{minipage}[t]{75mm}
%\begin{figure}
%\begin{center}
%%\epsfverbosetrue
%\leavevmode
%\epsfxsize=0.80\textwidth
%\epsfbox{vsshape.eps}
%\end{center}
%\caption{The vortex shell in $3+1$ dimensional space. }
%\label{fig:vsshape}
%\end{figure}
%\end{minipage}

 The motion of CDW is expressed by the extracted macroscopic 
degrees of freedom in these work but the dissipation 
due to the interaction between them and the other degrees of freedom
 is neglected, which can affect the conductance 
due to quantum tunneling \cite{CaLe83a,Weis94a}, and is important in 
comparison with experiments. 
 In this paper we investigate the effects of the ohmic dissipation on
the quantum phase slippage according to Caldeira--Leggett theory 
\cite{CaLe83a,Weis94a}.
 We write down the ``bounce'' action in \SECT{sec:form} in 
the case without dissipation, and add to this
the dissipation term coming from the interaction between the extracted 
macroscopic degree of freedom and the normal electrons in \SECT{sec:diss}. 
 We see that the evaluated dissipation term can modify
the dependence of the conductance in the electric field.
In \SECT{sec:conc}, we discuss the result obtained 
in \SECT{sec:diss} with that of experiments. 
 It turns out that such a dissipation can be neglected in experiments.

%%%%%%%%%%%%%%%%%%%%%%%%%%%%%%%%%%%%%%%%%%%%%%%
%\newpage
%\baselineskip=24pt
%%%%%%%%%%%%%%%%%%%%%%%%%%%%%%%%%%%%%%%%%%%%%%%%%%%%%%%%%%%
\section{Formalism}
\label{sec:form}
%%%%%%%%%%%%%%%%%%%%%%%%%%%%%%%%%%%%%%%%%%%%%%%%%%%%%%%%%%%

 We start from the 
Euclidean action of the FLR model
% Fukuyama--Lee--Rice model 
\cite{Fuku76a,FuLe78a,LeRi79a,MaVi90a}, which is given as follows, 
\begin{eqnarray}
S_{\rm E} &=& \int_{- \infty}^{\infty}  
{\rm d} x  {\rm d} y  {\rm d} z  {\rm d} \tau \frac{1}{4} N_0 f
\hbar \Bigg\{
\left[ \frac{m^\ast}{m} \left( \frac{\partial \phi}{\partial \tau} 
\right)^2 +v_{\rm x}^2 \left( \frac{\partial \phi}{\partial x} \right)^2
+v_{\rm y}^2 \left( \frac{\partial \phi}{\partial y} \right)^2
+v_{\rm F}^2 \left( \frac{\partial \phi}{\partial z} \right)^2 
\right] \nonumber \\
& & {} + \frac{4 e v_{\rm F}}{\hbar} \varepsilon \phi \Bigg\} + 
\int_{- \infty}^{\infty}  {\rm d} x  {\rm d} y  {\rm d} z  
{\rm d} \tau V_{\rm imp} \BK{\BM{r}}
\label{eq:MakiHAMa} \\
{\rm d} \tau \rho_1 \sum_j 
&=& \frac{N_0 f \hbar}{4} \frac{v_{\rm x} v_{\rm y}}{v_{\rm F}} 
\sqrt{\frac{m^\ast}{m}} \int_{- \infty}^{\infty}
 {\rm d} x'  {\rm d} y'  {\rm d} z' {\rm d} \tau'
 \Bigg\{ \left[ \left( \frac{\partial \phi}{\partial \tau'}
\right)^2 +\left( \frac{\partial \phi}{\partial x'} \right)^2
+\left( \frac{\partial \phi}{\partial y'} \right)^2
+\left( \frac{\partial \phi}{\partial z'} \right)^2 \right]
\nonumber \\
& & {} + \frac{4 e}{\hbar v_{\rm F}} \varepsilon \phi 
+ \frac{4}{N_0 f \hbar v_{\rm F}^2} V_{\rm imp} \BK{\BM{r}'} \Bigg\} , 
\label{eq:MakiHAMb}
\end{eqnarray}
where we choose $z$-axis as the one--dimensional direction. Here $\phi$ is 
the phase of CDW, $N_0 = \BK{\pi v_{\rm F} l_{\rm x} l_{\rm y}}^{-1}$, 
$m^{\ast}$ effective mass, $\varepsilon$ the applied electric field 
and $f$ is the condensation density and $f \sim 2$ 
in the low temperature range.
$(v_{\rm x}, v_{\rm y}, v_{\rm F})$ stand for the elements of Fermi 
velocity, and are given by 
$(\sqrt{2} t_{\rm x} l_{\rm x} / \hbar, \sqrt{2} t_{\rm y} l_{\rm y} /
\hbar, 2 t_{\rm z} l_{\rm z} \sin (l_{\rm z} k_{\rm F}) / \hbar)$, 
where $t_{\rm x},t_{\rm y},t_{\rm z}$ and 
$l_{\rm x},l_{\rm y},l_{\rm z}$ are 
transfer integrals and lattice constants in each direction, respectively. 
$V_{\rm imp} \BK{\BM{r}}$ is the impurity potential energy density. 
We introduce 
$x'=(v_{\rm F}/v_{\rm x})x, y'=(v_{\rm F}/v_{\rm y})y, z'=z$, and 
$\tau'=v_{\rm F} \tau \sqrt{m/m^\ast}$ and obtain 
the isotropic form, \SIKI{eq:MakiHAMb}.

 In order to calculate the tunneling probability, 
we must specify the state of the system just before the tunneling.  
 Duan considered the homogeneous CDW \cite{Duan93a}, 
which means there is no macroscopic deformation in the CDW.
 In this case, however, the estimated value of $\varepsilon_0$ 
in \SIKI{eq:exdep} is too large.
Maki considered the CDW pinned by strong pinning centers, which are, 
at first, assumed to be electric probes 
at both ends of the sample \cite{Maki95a}.
We put $\delta \phi \equiv \phi - \phi_{\rm m}$, where 
$\phi_{\rm m}$ is the phase of the metastable state, that is, the 
phase configuration before the tunneling. Then we have 
\begin{eqnarray}
\delta S_{\rm E} &=& \delta S_{\rm p}^{\BK{0}} + \delta S_{\rm p}^{\BK{1}}
+ \delta S_{\varepsilon {\rm field}} \label{eq:ea}\\
\delta S_{\rm p}^{\BK{0}} &=& 
A_0 \int  {\rm d}^4 r' \left( \partial_\mu {\it \delta} \phi \right)^2 
\label{eq:eap0}\\
\delta S_{\rm p}^{\BK{1}} &=& A_0 \int  {\rm d}^4 r' 
\left\{ 2 \left[ \partial_\mu \left( {\it \delta} \phi \right) \right]
 \cdot \partial_\mu \phi_{\rm m} \right\} \label{eq:eap1}\\
\delta S_{\varepsilon \ {\rm field}} &=& A_0 \int  {\rm d}^4 r' 
\frac{4 e}{\hbar v_{\rm F}} \varepsilon \delta \phi , \label{eq:eaef}
\end{eqnarray}
where 
$A_0 = N_0 f \hbar v_{\rm x} v_{\rm y} \sqrt{m^\ast /m} /4 v_{\rm F}$, 
$\mu = (\tau', x', y', z')$, and $\delta S_{\rm p}^{\BK{0}}$ 
represents the elastic and kinetic energy of the CDW in the case of 
vanishing $\phi_{\rm m}$, and 
$\delta S_{\rm p}^{\BK{1}}$ the contribution from the deformation 
of the CDW in the metastable state, 
and $\delta S_{\varepsilon \ {\rm field}}$ is the contribution 
from electric field.

 In the case of the homogeneous CDW, $\delta S_{\rm p}^{(1)}$ is 
negligible.
 However, in the case of the CDW pinned by strong pinning centers, 
CDW deforms so large, which is expressed between two strong pinning 
centers as,
\begin{eqnarray}
\phi_{\rm m} = \frac{e\varepsilon}{\hbar v_{\rm F}} z (z-L) ,
\label{eq:deform}
\end{eqnarray}
where $L$ is the distance between strong pinning centers, 
one of which locates at $z = 0$. In the above,  
we neglect the effect of impurities in the bulk, which is taken 
into account by slight modification of the amplitude 
of \SIKI{eq:deform} \cite{MaTa86}. Then \SIKI{eq:eap1} becomes
\begin{eqnarray}
\delta S_{\rm p}^{\BK{1}} &\simeq& - 2 A_0 
\ABS{\PDEL{\phi_{\rm m}}{z'}}_{z=z_0}
\int  {\rm d}x'  {\rm d}y'  {\rm d}\tau' \delta \phi, 
\label{eq:eap1a}
\end{eqnarray} 
where we note the contribution from one strong pinning center.
%where $L$ and $z_0$ are the distance between the strong pinning centers, 
%and between the quantum nucleation point and 
%the nearer strong pinning center, respectively.
Here it is assumed that $\delta \phi$ varies sharply only 
in the vicinity of $z=z_0$.
 \SIKIL{eq:eap1a} gives much larger contribution to the Euclidian 
action than \SIKI{eq:eaef} in a realistic case, 
so we can neglect the contribution which 
comes from $\delta S_{\varepsilon {\rm field}}$ \cite{Maki95a}, 
only which is taken into account in ref. $11$. %\cite{Duan93a}. 

 We can express the above action by employing the radius of the 
``vortex ring'' at imaginary time $\tau'$, $R \BK{\tau'}$, as follows, 
\begin{eqnarray}
\delta S_{\rm E} &=& 16 \pi^2 A_0 \int_0^{\tau_{\rm m}'} 
 {\rm d} \tau' R \BK{\tau'} \sqrt{1 + \BK{\WDEL{R \BK{\tau'}}{\tau'}}^2}
 \ln \frac{\tau_{\rm m}'}{\xi_z} \nonumber \\
 & & {} - \frac{4 \pi e A_0 
\BK{L + 2 \xi_{\rm z}}}{\hbar v_{\rm F}}
\varepsilon \times 2 \pi \int_0^{\tau_{\rm m}}  {\rm d} \tau' R \BK{\tau'}^2 ,
\label{eq:geneaction}
\end{eqnarray}
where 
$\tau_{\rm m}'$ is positive zero point of $R \BK{\tau'}$. 
 In the case without dissipation, ``vortex shell'' is considered 
to have a sphere shape with a constant radius $R$. In this case 
$\delta S_{\rm E}$ is given as 
\begin{eqnarray}
\delta S_{\rm E} = A R^2 - E R^3 , 
\label{eq:actionnd}
\end{eqnarray}
where
\begin{eqnarray}
A &=& 16 \pi^2 A_0 \ln \frac{R}{\xi_{\rm z}} \label{eq:ndconst1} , \\
E &=& \frac{16 \pi^2 e A_0 \BK{L + 2 \xi_{\rm z}}}{3 \hbar v_{\rm F}}
\varepsilon .
\label{eq:ndconst2}
\end{eqnarray}
We optimize $\delta S_{\rm E}$ about $R$, and get 
the bounce action, 
\begin{eqnarray}
S_{\rm B} = \frac{4 A^3}{27 E^2} .
\label{eq:bactionnd}
\end{eqnarray}
Then we have the conductance,
\begin{eqnarray}
\sigma \propto \exp{\left[ - \BK{\varepsilon_0 / \varepsilon}^2 
\right]}, \hspace{20pt} \varepsilon_0 = 
\frac{4}{e \BK{L + 2 \xi_{\rm z}}} 
\sqrt{\frac{2 \pi f t_{\rm x} t_{\rm y}}{3}}
\BK{\frac{m^\ast}{m}}^{\frac{1}{4}} \BK{\ln \frac{R^{\rm (e)}}
{\xi_{\rm z}}}^3 .
\label{eq:ndcond}
\end{eqnarray}
Here superscript (e) means extremum solution.
This form is the same as that observed in the experiment \cite{ZaZo93a}.
 In the case of the pure system and the two-probe method, 
$L$ is the distance between the probes, and then the value 
of $\varepsilon_0$ is much smaller than that of the experimental result.

 The action we have evaluated just above does not include 
the degrees of freedom of the normal electrons though the motion of 
the vortex ring accompanies the conversion from 
the condensed electrons to the normal electrons or opposite. 
So we take into consideration the interaction 
between the normal electrons and the CDW, which causes the dissipation. 
The latter is described by the macroscopic degree of freedom, 
that is the phase and the Peierls gap in the present case.

%%%%%%%%%%%%%%%%%%%%%%%%%%%%%%%%%%%%%%%%%%%%%%%%%%%%%%%%%%
%\newpage
%\baselineskip=24pt
%%%%%%%%%%%%%%%%%%%%%%%%%%%%%%%%%%%%%%%%%%%%%%%%%%%%%%%%%%%
\section{Effects of Dissipation}
\label{sec:diss}
%%%%%%%%%%%%%%%%%%%%%%%%%%%%%%%%%%%%%%%%%%%%%%%%%%%%%%%%%%%
 Now, we investigate effects of the dissipation. The 
motion of the ``vortex ring'' accompanies the conversion of 
the normal electrons to the condensed ones or opposite. 
It is considered that there are many normal electrons 
in the vicinity and inside of the vortex core even 
in semiconducting CDW system, such as TaS$_3$. It is 
difficult, however, to deal with the interaction between the CDW and 
the normal electrons microscopically \cite{Haya98a}. 
Here we consider the dissipation 
phenomenologically in accordance with the 
Cardeira--Leggett theory \cite{CaLe83a,Weis94a}. 
In this paper, we assume the ohmic 
dissipation which comes from the two types of interactions.
One is caused by the interaction between the CDW phase and 
the normal electrons in the vicinity of the vortex core. 
The other is due to the interaction between the CDW
and the normal electrons in the vortex core.
First we consider the former. 
The interaction is expressed by the CDW phase around the ``vortex ring'',
$\delta \phi$, as follows,
\begin{eqnarray}
\delta S_{\rm d}^{\rm (p)} [\delta \phi] = \frac{\eta \gamma^2}{2 \pi} 
\int  {\rm d}x'  {\rm d}y'  {\rm d}z'  {\rm d}\tau'  {\rm d}\tau''
\frac{\left[ \delta \phi \left(x',y',z',\tau' \right) - \delta \phi 
\left(x',y',z',\tau''\right) \right]^2}
{ \left( \tau' - \tau'' \right)^2}.
\end{eqnarray}
This action gives the frictional term $\eta \delta \dot{\phi}$ 
in the equation of motion of $\phi$.
Here $\gamma = \sqrt{\xi_{\rm x} \xi_{\rm y}} / \xi_{\rm z}$, 
$r' = \sqrt{x^{\prime 2} + y^{\prime 2}}$ 
and $\delta \phi \BK{x',y',z',\tau'}$ is given by
\begin{eqnarray}
\delta \phi \BK{x',y',z', \tau'} = -\frac{\pi}{2} 
- \arctan \gamma \frac{r' - R\BK{\tau'}}{z'} , \label{eq:voring}
\end{eqnarray}
when $\gamma R\BK{\tau}$ is the radius of the ``vortex ring'' at 
imaginary time $\tau$.
 We execute the partial integral about $\tau'$ and $\tau''$ and get 
\begin{eqnarray}
\delta S_{\rm d}^{\rm (p)} [\delta \phi] &=& - 2 \eta \gamma^4 
\int  {\rm d} \tau'  {\rm d} \tau''  {\rm d} r'  {\rm d} z' 
\frac{r' z^{\prime 2}}{\left\{ z^{\prime 2} + 
\left[ r'- R \BK{\tau'} \right]^2 \right\} 
\left\{ z^{\prime 2} + \left[ r'- R \BK{\tau''} \right]^2 \right\}} 
\nonumber \\
& & {} \times \PDEL{R \BK{\tau'}}{\tau'} \PDEL{R \BK{\tau''}}{\tau''} 
\ln \frac{\ABS{\tau' - \tau''}}{\xi_{\rm z}} 
\STEPF{\ABS{\tau' - \tau''} - \xi_{\rm z}}
, \label{eq:actionmi}
\end{eqnarray} 
where we introduce the step function to remove the self--interaction 
of the ``vortex ring''.
 The part that satisfies $\tau' + \tau'' =0$ 
gives large contribution to the integral because 
$R \BK{\tau} = R \BK{-\tau}$ and the part that satisfies 
$\tau' - \tau'' =0$ is excluded due to the step function 
$\STEPF{\ABS{\tau' - \tau''} - \xi_{\rm z}}$. We integrate 
eq.(\ref{eq:actionmi}) with respect to $r'$ and $z'$, but these integrals 
are complicated because their ranges depend on the sample size. 
Here, we assume one of the upper limits of the integral variables, 
$r'$ or $z'$, is so large and can be set infinity.
Then $\delta S_{\rm d}^{\rm (p)} [ \delta \phi ]$ is evaluated as, 
\begin{eqnarray}
\delta S_{\rm d}^{\rm (p)} [\delta \phi] &=& - 2 \eta \gamma^4 
\int_{ - \infty}^{\infty} {\rm d} \tau'  {\rm d} \tau'' 
\left[ L_{\Lambda} + C R \BK{\tau'} \right] 
\PDEL{R \BK{\tau'}}{\tau'} \PDEL{R \BK{\tau''}}{\tau''} \nonumber \\ 
 & & {} \times \ln \frac{\ABS{\tau' - \tau''}}{\xi_{\rm z}} 
\STEPF{\ABS{\tau' - \tau''} - \xi_{\rm z}}. 
\label{eq:actiontau}
\end{eqnarray}
When the upper limit of the integration about $z'$ is large enough,
$L_{\Lambda}$ and $C$ are given as follows,
\begin{eqnarray}
L_{\Lambda} \sim \frac{\pi l_{\rm s}}{2 \gamma}, \ \ 
C \sim \ln \frac{l_{\rm s}}{\xi_{\rm z}}, \label{eq:constls}
\end{eqnarray}
where $l_{\rm s}$ is the radius of the section of the sample. 
On the other hand, if that about $r'$ is large enough, we have
\begin{eqnarray}
L_{\Lambda} \sim \frac{L}{2 \gamma}, \ \
C \sim \frac{\pi}{4 \gamma} \ln \frac{L}{\xi_{\rm z}}. \label{eq:constlz}
\end{eqnarray}

 Next, we consider the latter interaction, that is, the interaction 
between the CDW and the normal electrons in the vortex core. 
In the vortex core, 
the gap vanishes and the Fermi surface is recovered. 
Since we cannot define the phase in the vortex core, it is 
necessary to describe the motion of the vortex ring in terms 
of the degree of freedom except for the phase. 
Here we can describe it in terms of its radius.
Then the dissipation term which derives the friction term 
proportional to $2 \pi R \BK{\tau'}  {\rm d} R \BK{\tau'} /  {\rm d} \tau'$ 
in the classical limit plays an important role.
 We can raise the following term for instance,
\begin{eqnarray}
\delta S_{\rm d}^{\rm (v)} [\delta \phi] = 
\frac{\eta_{\rm v} \gamma^3}{4 \pi} 
\int  {\rm d}\tau'  {\rm d}\tau'' 2 \pi R \left(\tau' \right)
\frac{\left[ R \left(\tau' \right) - R 
\left(\tau''\right) \right]^2}
{ \left( \tau' - \tau'' \right)^2}, 
\end{eqnarray}
where $\eta_{\rm v}$ is the viscosity coefficient per unit length. 
In fact this term gives the above friction term
in the equation of motion of $R \BK{\tau}$.
 We execute the partial integral about $\tau$ and $\tau'$, and obtain 
\begin{eqnarray}
\delta S_{\rm d}^{\rm (v)} [\delta \phi] &=& 
- \frac{\eta_{\rm v} \gamma^3}{2}
\int_{ - \infty}^{\infty}  {\rm d} \tau'  {\rm d} \tau'' 
R \BK{\tau'} \PDEL{R \BK{\tau'}}{\tau'} \PDEL{R \BK{\tau''}}{\tau''} 
\nonumber \\
 & & \times \ln \frac{\ABS{\tau' - \tau''}}{\xi_{\rm z}} 
\STEPF{\ABS{\tau' - \tau''} - \xi_{\rm z}}.
\end{eqnarray}

 Thus, the total dissipation term of the Euclidean action is
\begin{eqnarray}
\delta S_{\rm d} [\delta \phi] &=& 
- \int_{ - \infty}^{\infty} d \tau' d \tau''
\BK{2 \eta \gamma^4 L_{\Lambda} + 2 \eta \gamma^4 C R \BK{\tau'} 
+ \frac{\eta_{\rm v} \gamma^3}{2} R \BK{\tau'}}
\PDEL{R \BK{\tau'}}{\tau'} \PDEL{R \BK{\tau''}}{\tau''} \nonumber \\
 & & {} \times \ln \frac{\ABS{\tau' - \tau''}}{\xi_{\rm z}} 
\STEPF{\ABS{\tau' - \tau''} - \xi_{\rm z}}.
\end{eqnarray}

In principle we can get the ``bounce'' solution by variating the whole
action with respect to $R\BK{\tau}$.
It is, however, difficult in general.
 So, we evaluate the above action in the weak and the
strong damping regimes \cite{Hida84a,Hida85a}.
 In the weak damping regime, we can approximate the shape of the 
``vortex shell'' by a sphere.
Then the dissipation term derives the term which is proportional to
the surface of the ``vortex shell'' and that proportional to 
its volume. Therefore, $A$ and $E$ in \SIKI{eq:actionnd} is 
expressed as
\begin{eqnarray}
   A &=& 16 \pi^2 A_0 \ln \frac{\tau_{\rm m}^{{\rm (e)}\prime} }{\xi_{\rm z}}
+ \pi^2 \eta \gamma^4 L_{\Lambda} 
\ln \frac{2 \tau_{\rm m}^{{\rm (e)}\prime} }{\xi_{\rm z}},
\label{eq:shift1} \\
   E &=& 4 \pi e A_0 \BK{L + 2\xi_{\rm z}}\varepsilon /\hbar v_{\rm F} 
- \pi \eta \gamma^4 C 
\ln \frac{2 \tau_{\rm m}^{{\rm (e)}\prime} }{\xi_{\rm z}}
- \frac{\pi}{4}\eta_{\rm v} \gamma^3 
\ln \frac{2 \tau_{\rm m}^{{\rm (e)}\prime} }{\xi_{\rm z}}
.
\label{eq:shift2}
\end{eqnarray}
Then the conductance is expressed as
\begin{eqnarray}
\sigma \propto \exp{\left[ - S_{\rm B} / \hbar \right]}, 
\hspace{20pt} S_{\rm B} / \hbar =
\BK{\frac{\tilde{\varepsilon}_0}{\varepsilon - \varepsilon_1}}^2 
,\label{eq:wdcond1}
\end{eqnarray}
where
\begin{eqnarray}
\tilde{\varepsilon}_0 &=&
  \BK{1 + 
    \frac{
      \eta \gamma^4 L_{\Lambda} \ln \frac{
        2 \tau_{\rm m}^{{\rm (e)}\prime}
      }{
        \xi_{\rm z}
      }
    }{
      16 A_0 \ln \frac{
        \tau_{\rm m}^{{\rm (e)}\prime}
      }{
        \xi_{\rm z}
      }
    }
  }^3 \varepsilon_0 , \\
\varepsilon_1 &=& 
    \frac{
      \eta \gamma^4 C \hbar v_{\rm F} \ln \frac{
        2 \tau_{\rm m}^{{\rm (e)}\prime}
      }{
        \xi_{\rm z}
      }
    }{
      4 e A_0 \BK{L + 2\xi_{\rm z}} 
    }
+ \frac{
     \eta_{\rm v} \gamma^3 \hbar v_{\rm F} \ln 
     \frac{\tau_{\rm m}^{{\rm (e)}\prime}}{\xi_{\rm z}}
  }{
    16 e A_0 \BK{L + 2 \xi_{\rm z}}
  }
.\label{eq:e0e1}
\end{eqnarray}
The condition that the assumption of the weak damping is valid is 
expressed as, 
\begin{eqnarray}
\eta \gamma^4 L_{\Lambda} \ll 16 A_0 \hspace{12pt}\mbox{and}\hspace{12pt}
\varepsilon \gg \varepsilon_1
\label{eq:wdconda}.
\end{eqnarray}
 Comparisons of \SIKI{eq:ndcond} and \SIKI{eq:wdcond1} indicates 
that the dissipation increases $\varepsilon_0$ in \SIKI{eq:bactionnd} 
and decreases the applied field $\varepsilon$.
 Thus, the tunneling probability is suppressed in two ways.
It is to be noted that in the range $\varepsilon \simeq \varepsilon_1$ 
the assumption of the weak damping and then \SIKI{eq:wdcond1} are not 
valid. 

 In the case that $\eta \gamma^4 L_{\Lambda} \gg 16 A_0$ and 
$\varepsilon \ll \varepsilon_1$, 
the system is in the strong damping region.
 In this region, the action becomes small 
when the part of $R\BK{\tau'}$ parallel to $\tau'$ axis 
becomes large \cite{Hida84a}.
 Hence, we can assume the shape of the ``vortex shell'' 
by the cylinder. Then we have
\begin{eqnarray}
\delta S_{\rm E} &=& 16\pi^2 A_0 \BK{\frac{1}{2} R_{\rm d}^2 
+ 2 R_{\rm d} T_{\rm d}} \ln \frac{T_{\rm d}}{\xi_{\rm z}} 
- \frac{8 \pi^2 e A_0 \BK{L + 2 \xi_{\rm z}}}{\hbar v_{\rm F}} 
\varepsilon R_{\rm d}^2 T_{\rm d} \nonumber \\
 & & {} + \BK{4 \eta \gamma^4 L_{\Lambda} + 2 \eta \gamma^4 C R_{\rm d}
+ \frac{1}{2} \eta_{\rm v} \gamma^3 R_{\rm d} }
 R_{\rm d}^2 
\ln \frac{2 T_{\rm d}}{\xi_{\rm z}},
\end{eqnarray}
where $2T_{\rm d}$ is the height of the cylinder, 
and $R_{\rm d}$ is its radius. 
We optimize the action about $T_{\rm d}$ and $R_{\rm d}$, and get 
\begin{eqnarray}
R_{\rm d}^{\rm (e)} &=& \frac{4 \hbar v_{\rm F}}
{e \varepsilon \BK{L + 2 \xi_{\rm z}}} 
\ln \frac{T_{\rm d}^{\rm (e)}}{\xi_{\rm z}} , \\
T_{\rm d}^{\rm (e)} &=& 
\frac{
        \hbar v_{\rm F} \left[
          8 \pi^2 A_0 + 4 \eta \gamma^4 L_\Lambda + \BK{
            2 \eta \gamma C + \eta_{\rm v}
          } \gamma^3 R_{\rm d}^{\rm (e)}
        \right]
     }{
       4 \pi^2 e A_0 \BK{L + 2 \xi_{\rm z}} \varepsilon
     }
\ln \frac{T_{\rm d}^{\rm (e)}}{\xi_{\rm z}}
.\label{eq:sdtc}
\end{eqnarray}
Then the conductance is given as,
\begin{eqnarray}
\sigma \propto \exp \left\{ - \left[ 
    \BK{\frac{\tilde{\varepsilon}_0^\prime}{\varepsilon}}^2 
  + \BK{\frac{\tilde{\varepsilon}_2^\prime}{\varepsilon}}^3 
  \right] \right\}
,
\end{eqnarray}
where
\begin{eqnarray}
\tilde{\varepsilon}_0^\prime &=& 
  \frac{
    4 v_{\rm F} \sqrt{
      \hbar \BK{
        8 \pi^2 A_0 + 4 \eta \gamma^4 L_\Lambda
      } 
    }
  }{
    e \BK{L + 2 \xi_{\rm z}}
  }
\ln \frac{T_{\rm d}^{\rm (e)}}{\xi_{\rm z}}
\label{eq:stdcond1}
\\
\tilde{\varepsilon}_2^\prime &=& 
  \frac{
    4 \hbar^{\frac{2}{3}} v_{\rm F} \gamma \left[
      \BK{
        2 \eta \gamma C + \eta_{\rm v}
      } \ln \frac{T_{\rm d}^{\rm (e)}}{\xi_{\rm z}}
    \right]^{\frac{1}{3}}
  }{
    e \BK{L + 2 \xi_{\rm z}}
  }
. \label{eq:stdcond}
\end{eqnarray}
 Thus the logarithm of the conductance can be proportional to the
reciprocal of the electric field cubed.
 As seen from \SIKI{eq:wdcond1} and \SIKI{eq:stdcond1}, there exists 
the electric field $\varepsilon^\ast$ crosses from the strong damping 
to the weak damping regimes.
 The magnitude of $\varepsilon^\ast$ is the same order with 
that of $\varepsilon_1$.

%%%%%%%%%%%%%%
%\newpage
%\baselineskip=24pt
%%%%%%%%%%%%%%%%%%%%%%%%%%%%%%%%%%%%%%%%%%%%%%%%%%%%%%%%%%%
\section{Concluding Remarks}
\label{sec:conc}
%%%%%%%%%%%%%%%%%%%%%%%%%%%%%%%%%%%%%%%%%%%%%%%%%%%%%%%%%%%

In this section, we estimate the effects of the dissipation 
in experiments of the conductance.
The value of each parameter is obtained in refs. $6$, $14$, $21$.
%\cite{ZaZo93a}, \cite{Maki95a}, \cite{TuLyGa88}.

We first discuss the case that the dissipation comes from the
interaction between the CDW phase $\phi$ and the normal electrons 
in the vicinity of the vortex core.
 The coupling constant $\eta$ in this case is considered
to be proportional to the normal electron density near the vortex and 
then to be expressed as $\eta = g \rho_{\rm n}^{\rm v}$,
where $g$ is a coupling constant per electron and $\rho_{\rm n}^{\rm v}$ is 
the normal electron density near the vortex. 
 We estimate the effects in ${\rm NbSe}_3$ at first, where Fermi surface
survives even at low temperature and then $\rho_{\rm n}^{\rm v}$ 
is expected to have similar magnitude to that in the bulk, 
$\rho_{\rm n}^{\rm b}$.
 The magnitude of $g \rho_{\rm n}^{\rm b}$ can be evaluated from the bulk 
conductance under large electric field, where the dissipation 
for the sliding motion of CDW in this material mainly comes
from the interaction between the CDW phase and the normal electron 
\cite{FlCa86}.
Then we obtain
\begin{eqnarray}
\frac{\eta \gamma^4 L_{\Lambda}}{16 A_0} &\sim&
2 \times 10^{-7}
[{\rm m}^{-1}] \times L_{\Lambda}[{\rm m}] ,
\label{eq:suuchia} \\
\frac{\eta \gamma^4 C \hbar v_{\rm F}}{4 e A_0 \BK{L + 2 \xi_{\rm z}}
\varepsilon} &\sim& 3 \times 10^{-16}[{\rm V}]
\times
\frac{C}{\varepsilon \BK{L + 2 \xi_{\rm z}}}[{\rm V}^{-1}].
\label{eq:suuchib}
\end{eqnarray}
Since $L_{\Lambda}$ is $\pi l_{\rm s}/ 2 \gamma
(\sim 3 \times 10^{-6}[{\rm m}])$
or $\BK{L + 2 \xi_{\rm z}} / 4 \gamma^2 (\sim 2 \times
10^{-1}[{\rm m}])$, \SIKI{eq:suuchia} is much smaller than unity.
On the other hand, $C$ is the quantity of ${\cal O} \BK{1}$, and
\SIKI{eq:suuchib} is much smaller than unity, too.
Therefore, we conclude the effect of such a dissipation is 
negligible in ${\rm NbSe}_3$.
 Next we try to estimate the same effects in o-${\rm TaS}_3$.
In this case we do not know the magnitude of $\eta$.
It is expected, however, that it is similar to or smaller than that
in ${\rm NbSe}_3$.
 Hence the effect of such a dissipation is negligible even in this
material.

 Next we consider the case that the dissipation comes from the interaction
between the CDW and the normal electron in the vortex core.
 In this case the dissipation has quantitatively similar effect 
both in NbSe$_3$ and TaS$_3$. So we estimate it for NbSe$_3$.
 The motion of the vortex accompanies the interior 
normal electrons, which enhance the vortex mass and cause the dissipation.
 The enhancement of mass is discussed in Appendix. 
 The dissipation coefficient $\eta_{\rm v}$ is estimated as follows,
\begin{eqnarray}
\eta_{\rm v} \simeq \pi e^2 {\rho_{\rm n}^{\rm vc}}^2 \xi_{\rm z} 
\sqrt{\xi_{\rm x} \xi_{\rm y}} 
\rho_{\rm res}, \label{eq:etav}
\end{eqnarray}
where $\rho_{\rm res}$ is the residual resistivity, 
$\rho_{\rm n}^{\rm vc}$ is the electron density in the vortex core,
respectively.
 From \SIKI{eq:e0e1},
$\varepsilon_1$ is given as follows,
\begin{eqnarray}
\varepsilon_1 \sim 2 \times 10^{-7} [{\rm V}] / L [{\rm m}]
\label{eq:shifte}.
\end{eqnarray}
This value is much smaller than $\varepsilon_0 \simeq 
3 \times 10^{-1} [{\rm V}] / L [{\rm m}]$.
 Then we conclude that the dissipation effect can be neglected 
also in this case both for NbSe$_3$ and TaS$_3$. 

 In order to get the agreement of the magnitude of $\varepsilon_0$ 
with the experiment, we must set $L \simeq 3 \times 10^{-5} [{\rm m}]$, 
which is much smaller than the sample length 
$L_{\rm s} = 3.2 \times 10^{-4} [{\rm m}]$. 
 So we consider strong impurities in the sample may play the role of 
strong pinning centers.

 Before the end of this section, we would like to discuss the reason 
why the nonlinear conduction below $\varepsilon_{\rm T}$ is observed 
only in thin samples \cite{Mats95a}.
 In order to observe the stationary current the CDW in the bulk as well as 
that near the strong pinning centers such as probes must move.
Both motions are due to the quantum tunneling in the case discussed here.
The quantum tunneling in the bulk is considered to be due to 
the kink-antikink pair creation\cite{Maki77a}.
 The nucleation rate $\Gamma$ of kink-antikink pair in $d$ dimension
is given by
\begin{eqnarray}
\Gamma \propto{\rm d}^{- \BK{\epsilon_d / \varepsilon}^d} . \label{eq:kap}
\end{eqnarray}
The conductance of the system is proportional
to the product of the tunneling rate near the strong pinning center,
eg. \SIKI{eq:ndcond} and that in the bulk, \SIKI{eq:kap}.
When the sample is thin and its radius, $l_{\rm s}$, is smaller than 
the FLR coherence length in the perpendicular direction 
to the $1$--dimensional axis, $\xi_{{\rm x},{\rm y}}$, 
we can regard the system as 1--dimensional
in the bulk. It is to be noted that $\xi_{{\rm x},{\rm y}}$ is 
much larger than $R^{\rm (e)}$ and then the system can be 
$3$--dimensional for the the vortex ring.
Then the tunneling in the bulk is easier than that near the 
strong pinning centers, and the conductance
is determined by the latter, that is, the nucleation rate of 
the ``vortex ring''.
 However, when the sample becomes so thick that 
$l_{\rm s}$ is larger than $\xi_{{\rm x},{\rm y}}$, 
the system is regarded as
3--dimensional\cite{123d}.
Since $\Gamma$ is too small for $d = 3$, we can not observe tunneling 
current in this case.
 Hence we can conclude that the nonlinear conduction below 
$\varepsilon_{\rm T}$ is observed only in thin samples \cite{note1}.

 In this paper, we consider the effect of the dissipation on the CDW. 
 We can discuss it on the spin density wave (SDW) in the same way.
 Moreover, it is concerned with the quantum phase slippage
in superfluid $^4{\rm He}$ \cite{MuVi84},
and ultrathin superconducting wires \cite{Duan95a}.

%%%%%%%%%%%%%%%%%%%%%%%%%%%%%%%%%%%%%%%%%%%%%%%%%%%%%%%%%%%%%%%%%
%\acknowledgments
\section*{Acknowledgements}
 We would like to thank H. Fukuyama, K. Hida, and M. Hayashi 
for helpful discussions.
 This work was supported by the Research Fellowship of Japan Society 
for the Promotion of Science for Young Scientists, 
and Grants-in-Aid for Scientific Research of Ministry of Education, 
Science and Culture.
% We thank  for helpful conversations.
%%%%%%%%%%%%%%%%%%%%%%%%%%%%%%%%%%%%%%%%%%%%%%%%%%%%%%%%%%%%%%%%%
\appendix
\section{The Effective Mass of the Vortex Core}
\label{appendix:aa}

 We investigate the effective mass of the vortex core, $m_{\rm v}^{\ast}$.
The effective mass coming from \SIKI{eq:geneaction} is obtained as follows.
The term in the equation of motion with respect to $R\BK{\tau}$ 
from \SIKI{eq:geneaction} is given by,
\begin{eqnarray}
\frac{\delta S_{\rm E}}{\delta R\BK{\tau}} &=&
\BK{8 \pi^2 A_0 \ln \frac{\tau}{\xi_{\rm z}}} \left\{
\frac{1}{\sqrt{1+\BK{\PDEL{R\BK{\tau}}{\tau}}^2}}
- R\BK{\tau} \frac{\partial^2 R\BK{\tau}}{\partial \tau^2}
\left[1+\BK{\PDEL{R\BK{\tau}}{\tau}}^2 \right]^{- \frac{3}{2}}
\right\} \nonumber \\
 & & {} - \frac{16 \pi^2 e A_0 L \varepsilon}{\hbar} R\BK{\tau}
\label{eqn:moeqr} .
\end{eqnarray}
 We assume that
$\left| \partial R\BK{\tau}/\partial \tau \right| \ll 1$,
which shows that the velocity of vortex core is much smaller than
the phason velocity. We use this condition, and expand
\SIKI{eqn:moeqr}.
Then we get the effective mass per unit length
$m_{\rm v}^{\ast}$ from the coefficient of
$\partial^2 R\BK{\tau}/\partial \tau^2$ as follows,
\begin{eqnarray}
m_{\rm v}^\ast \simeq 4 \pi A_0 \ln 
\frac{\tau_{\rm m}^{\rm (e) \prime}}{\xi_{\rm z}}
\label{eqn:mvast} .
\end{eqnarray}
Next we take into account the contribution from the normal electrons 
in the vortex core to $m_{\rm v}^{\ast}$.
We add the following kinetic term to the Euclidian action,
\begin{eqnarray}
\delta S_{\rm vc} \simeq \gamma^3 v_{\rm F} \sqrt{\frac{m}{m^\ast}}
\int {\rm d} \tau' \pi^2  \rho_{\rm n}^{\rm vc} \xi_{\rm z}
\xi_{\rm T} R \BK{\tau'} \BK{\PDEL{R\BK{\tau'}}{\tau'}}^2 .
\label{eqn:elvc}
\end{eqnarray}
The effective mass coming from the above term is given as,
\begin{eqnarray}
m_{\rm vc}^\ast \simeq \pi \gamma^3 v_{\rm F} \sqrt{\frac{m}{m^\ast}}
\rho_{\rm n}^{\rm vc} \xi_{\rm z} \xi_{\rm T} .
\label{eqn:mvcast}
\end{eqnarray}
The comparison of
$m_{\rm v}^\ast$ and $m_{\rm vc}^\ast$
shows that
$m_{\rm vc}^\ast$ is always negligible.
Then we can take into account only the dissipation term and 
neglect others which come form the motion of the normal electrons 
in the vortex core.

%%%%%%%%%%%%%%%%%%%%%%%%%%%%%%%%%%%%%%%%%%%%%%%%%%%%%%%%%%%
% \newpage
%%%%%%%%%%%%%%%%%%%%%%%%%%%%%%%%%%%%%%%%%%%%%%%%%%%%%%%%%%%
% now the references. delete or change fake bibitem. delete next three
%   lines and directly read in your .bbl file if you use bibtex.

%\end{thebibliography}

%%%%%%%%%%%%%%%%%%%%%%%%%%%%%%%%%%%%%%%%%%%%%%%%%%%%%%%%%%%
% figures follow here

\newpage

%\begin{minipage}[t]{80mm}
%\ 
%\end{minipage}
%%
%\hspace{\fill}
%%
%\begin{minipage}[t]{75mm}
\begin{figure}
\begin{center}
%\epsfverbosetrue
\leavevmode
\epsfxsize=0.60\textwidth
\epsfbox{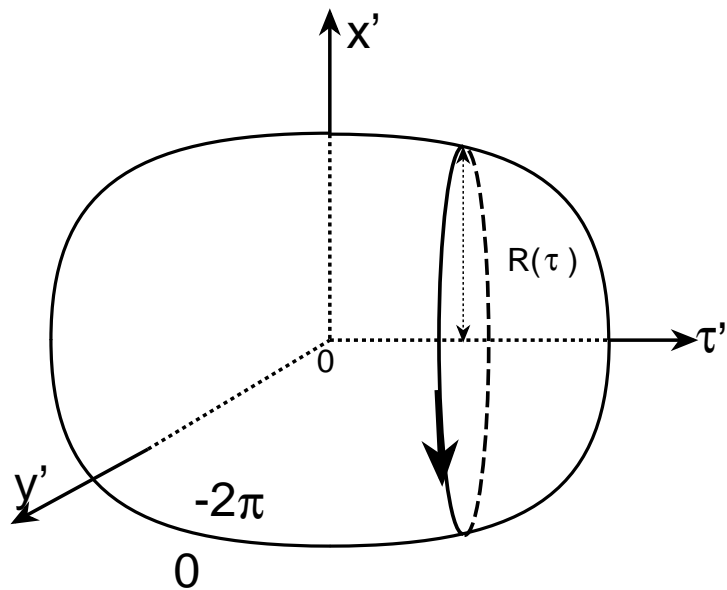}
\end{center}
\caption{The vortex shell in $3+1$ dimensional space. }
\label{fig:vsshape}
\end{figure}
%\end{minipage}

\begin{references}
\bibitem{123d} The system is considered to be 1 or 3--dimensional system 
in the bulk because it is isotropic in the plane perpendicular 
to the 1--dimensional axis.
\bibitem{note1} The system is rather isotropic in the plane perpendicular 
to the $1$--dimensional axis. So the system is $1$ or $3$--dimensional.
\end{references}
\end{document}